\documentclass{rspublic}
\usepackage{graphicx}
\usepackage{natbib}

\begin{document}

\title[Beyond crystals]{Beyond crystals: the dialectic of \\ materials and information}

\author[Julyan Cartwright and Alan Mackay]{Julyan H. E. Cartwright$^1$ and Alan L. Mackay$^2$}
\affiliation{
$^1$Instituto Andaluz de Ciencias de la Tierra, CSIC--Universidad de Granada, Campus Fuentenueva, E-18071 Granada, Spain \\
$^2$Birkbeck College, University of London, Malet Street, London WC1E 7HX, United Kingdom}

\label{firstpage}

\date{Version of \today}

\maketitle

\begin{abstract}{Crystallography; Information;Complexity; Materials; Self-assembly; Self-organization}
We argue for a convergence of crystallography, materials science, and biology, that will come about through asking materials questions about biology and biological questions about materials, illuminated by considerations of information. The complex structures now being studied in biology and produced in nanotechnology have outstripped the framework of classical crystallography and a variety of organizing concepts are now taking shape into a more modern and dynamic science of structure, form, and function. Absolute stability and equilibrium are replaced by metastable structures existing in a flux of energy carrying information and moving within an energy landscape of complex topology. Structures give place to processes and processes to systems. The fundamental level is that of atoms.  As smaller and smaller groups of atoms are used for their physical properties quantum effects become important;  already we see quantum computation taking shape. Concepts move towards those in life with the emergence of specifically informational structures. We now see the possibility of the artificial construction of a synthetic living system,  different from biological life, but having many or all of the same properties. Interactions are essentially nonlinear and collective. Structures begin to have an evolutionary history with episodes of symbiosis. Underlying all the structures are constraints of time and space. Through hierarchization, a more general principle than the periodicity of crystals, structures may be found within structures on different scales. We must integrate unifying concepts from dynamical systems and information theory to form a coherent language and science of shape and structure beyond crystals. To this end, we discuss the idea of categorizing structures based upon information according to the algorithmic complexity of their assembly.

\end{abstract}

\begin{quote}
Where is the wisdom we have lost in knowledge? \\
Where is the knowledge we have lost in information?  \\ (T. S. Eliot, The Rock) \\ \\
Where is the information we have lost in data? \\
Where are the data we have lost in noise?Ó \\ (our addition)
\end{quote}

\section{Introduction}

The historical objectives of crystallography were accomplished some years ago. It is now possible to find the arrangement of the atoms in any kind of ordered solid material. 
%The `phase problem'  has been solved, partly by using electron microscopy, where the phases of the scattered beams are not lost, and partly by `direct methods'â using relationships between the phases consequent on the knowledge that the electron density is everywhere non-negative. 
The subject of crystallography as an academic department --- as brilliantly put into practice by Bernal \citep{Brown} --- has not yet been replaced by a coherent new set of objectives, methods and materials. Now there are the areas of bioinformatics and of structural and molecular biology, but also the new area of nanotechnology, which has developed from materials science. How do we proceed from this point towards the integration of materials with information, both theoretically and experimentally, leading on to the imitation of living systems and the reproduction of their methods?

The International Union of Crystallography provides --- in our view --- a retrograde definition of a crystal as a structure giving a diffraction pattern with discrete points \citep{IUCr}. This definition characterizes a structure by a technique for observing it and imposes the characteristic limitations of the observer. At the conclusion of this paper we shall put forward a different means of categorizing structures, based on information. But let us begin with the physical conception of a crystal as a stable structure of atoms in a minimum energy configuration. This may be the absolute minimum, like the cubic structure of a sodium chloride crystal, so that there may be no options and the configuration thus carries no information (information being defined in information theory as the number of possible states of the system, or ``that which can distinguish one thing from another''). On the other hand, an arrangement of atoms, as for example a polytype of silicon carbide, seen as a stack of variously cubically or hexagonally close-packed layers, resembles a punched paper tape and such an aperiodic crystal may be seen to carry arbitrary information in the stacking sequence of the layers. 

The biological world was revolutionized by the realization that DNA has precisely such an aperiodic structure, that the sequence of bases in DNA carries explicit or intentional information (i.e., meaning; or what Eliot above called knowledge) via the genetic code, and that this sequence controls, via RNA and ribosomes, the structure of proteins and is essential to the processes of life. In these processes (in metabolism, growth, morphogenesis, and evolution) structure (the arrangement of atoms) and information (here seen as the sequence of bases in DNA, a particular feature of the arrangement of atoms) are seen to interact in systems through which energy flows. Living organisms are thus seen to have structures where information is concentrated, but both DNA and proteins are made of the same atoms following the same laws of chemistry, all based in turn upon physics.

 But information may be seen also to be distributed in all structures. In biology we find hierarchies of structure upon structure from the atomic to the human scale.  In chemistry, materials science, and nanotechnology we see the syntheses of increasingly complex structures chosen by the synthesizers. How can we characterize their complexity? What implicit or explicit information do they contain?  In many cases we find an energy landscape, or configuration space, of great complexity with many local energy minima connected by various paths.

In parallel with these developments, the world of computing has grown up, explicitly storing and processing information in material form, as the arrangement of atoms, or their quantum states. A correspondence should be borne in mind between the ways in which a computer sorts data and the ways in which atoms sort themselves out while proceeding from a vapour, a melt, or a solution to a solid structure. 

We thus have three parallel worlds: the biological (living processes and their manipulation and synthetic development), the inorganic (materials and their controlled synthesis), and the informatic (the computer world and information processing, including the necessary mathematics and algorithms for this end). In each of these worlds we find the partial segregation of information-bearing structures.

In this issue of Philosophical Transactions we have brought together a group of people who are thinking deeply about these questions. Our aim in presenting the papers in this issue is to spur on the development of this science of shape and structure beyond crystals.

\begin{figure}[tp]
\centering\includegraphics*[width=\textwidth,clip=true]{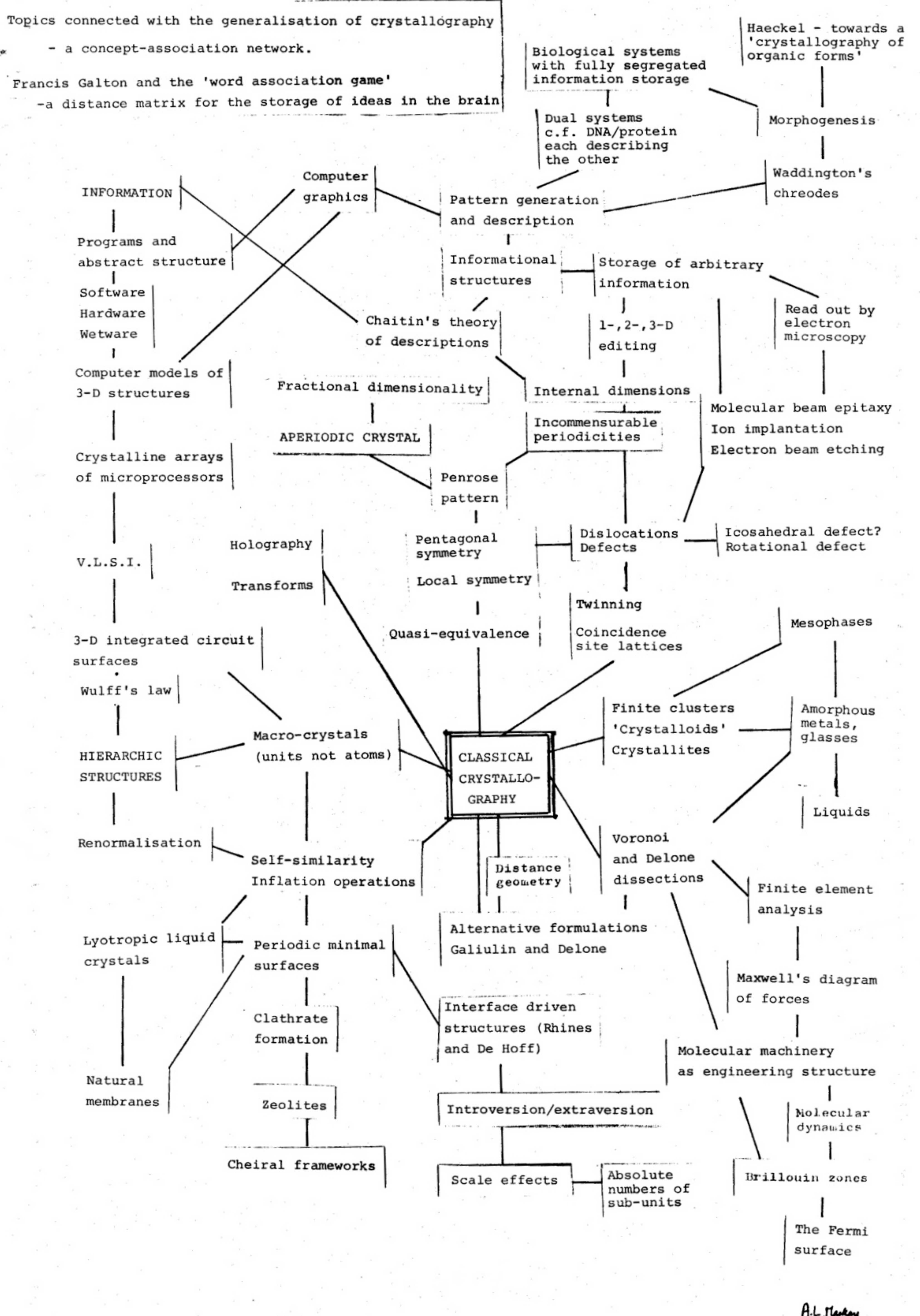} 
\caption{An association diagram for the generalization of crystallography developed by one of us (ALM) in 1982.
\label{alan1}}
\end{figure}

\section{Structure and information --- beyond classical crystallography}

How do structure and information interact? To resolve this question we must bring concepts of information theory to bear on crystallography and materials science and we must move from thinking only of static structures to considering dynamic patterns.  Classical crystallography deals with equilibrium structures, with the ÔabsoluteÕ identity of units and of their surroundings. There is long-range order, leading to the 230 space groups. The insistence on equilibrium means that how the structure is arrived at --- crystal growth --- is often not considered to be crystallography `proper', so that the dynamical processes that lead to the static crystal structure are ignored by crystallographers.

\subsection{Generalized crystallography}

Ideas on generalizing crystallography have been incubating for a long time. Bernal discussed them in the 1960s \citep{Bernal1967, Bernal1968}, and one of us (ALM) wrote about them from the 1970s on \citep{alan1975,alan1986,alan1995,alan2002}; the association diagram on generalized crystallography shown in Fig.~\ref{alan1} was produced  in 1982. Today we see areas of research where there are structures, more general than crystals, which are emerging from materials work and that are approaching biology,
in the sense that some are to be found in living materials. Likewise there are biological structures that are beginning to be understood in materials terms. Which are the threads connecting abiotic and biological structures and systems?
What prospects are there for constructing new structures?

\begin{figure}[tp]
\centering\includegraphics*[width=\textwidth,clip=true]{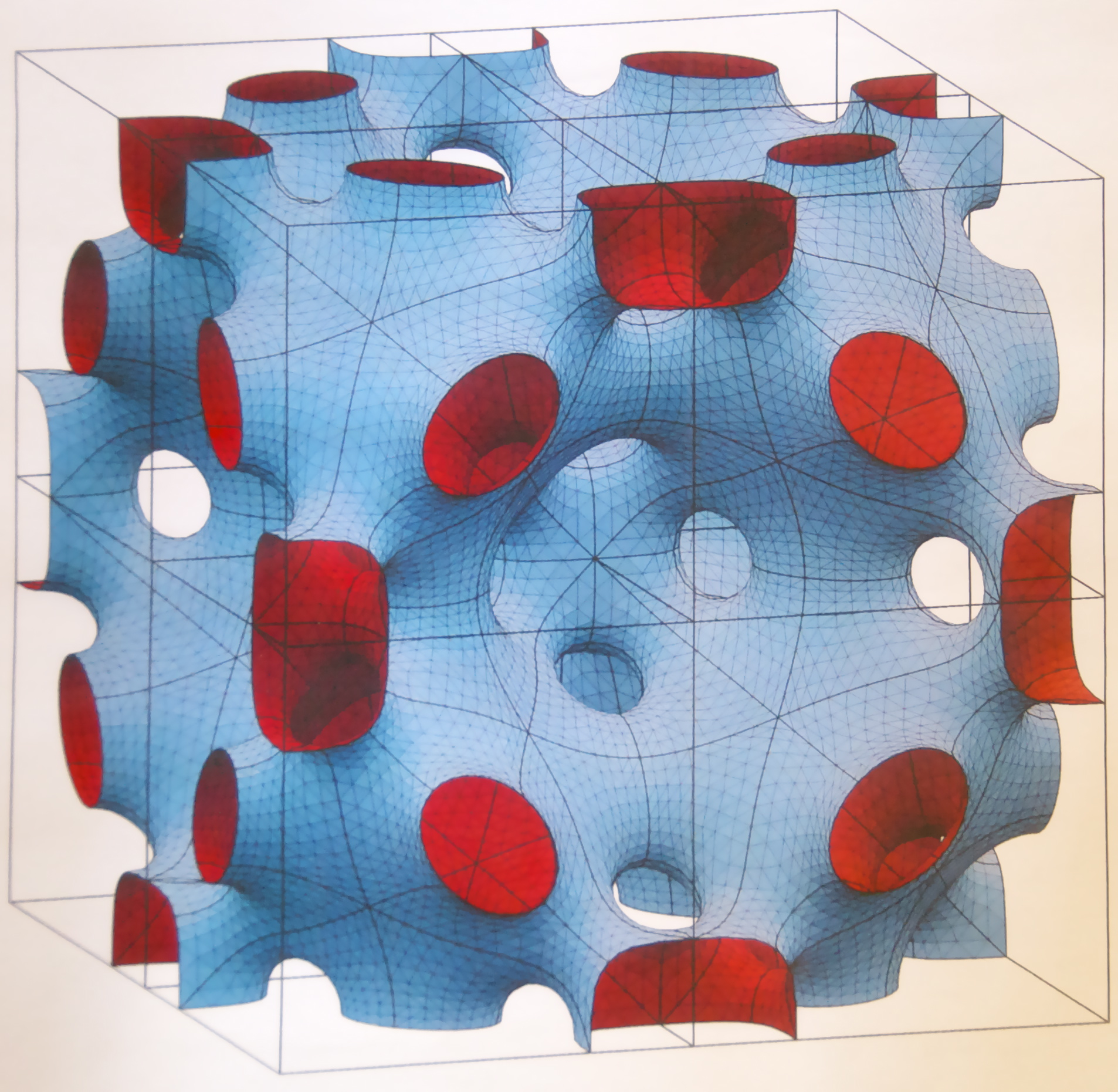} 
\caption{Cubic periodic minimal surface by Sho Takahashi.
\label{takahashi1}}
\end{figure}

\begin{figure}[tp]
\centering\includegraphics*[width=\textwidth,clip=true]{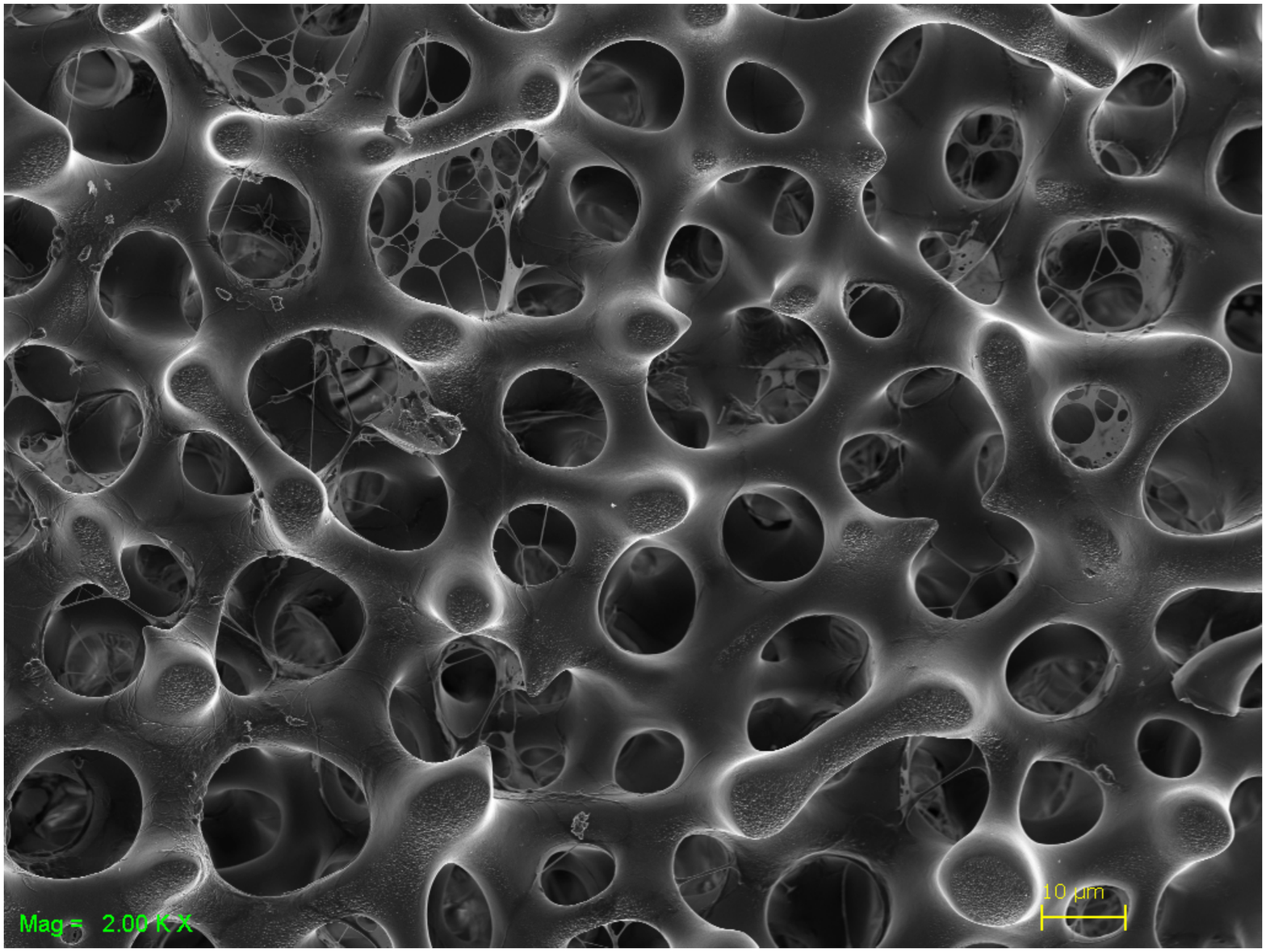} 
\caption{Scanning electron micrograph of a sea urchin (\emph{Paracentrotus lividus}) skeleton; image courtesy of Antonio Checa.
\label{sea_urchin1}}
\end{figure}

Some of the simplest of these structures are the membranes and micelles that divide the world into inside and outside. Aggregates of atoms or molecules begin to have emergent properties like vapour pressure, surface tension, and so on.
Such collective properties can lead molecules to assemble into membranes. Membranes in turn have emergent properties
like curvature that are involved in biological processes like gastrulation. With such membranes and surfaces, the ratio of surface to volume may dictate new topologies \citep{lord}. The figures displayed on the cover of this issue and in Fig.~\ref{takahashi1} are pieces of art by Sho Takahashi that reproduce examples of periodic minimal surfaces \citep{alan1985,alan2003}. Such geometries, sometimes with and sometimes without periodicity, are common in biological systems; compare Fig.~\ref{sea_urchin1}. One question is to understand how they are produced. And an open challenge for materials science is to be able to reproduce such surfaces at a molecular level in a material such as graphene \citep{alan1991,Mackay1993,Terrones2004}.

Membranes and micelles are but one manifestation of liquid and colloidal crystalline structures. Liquid-crystal geometries are more varied than those of solid crystals owing to the greater possibilities for attractive and repulsive interactions of units of a given shape in a fluid medium. Liquid crystals, although first discovered in biological media (hence the name \emph{cholesteric} for a common type of liquid crystalline order first found in cholesterol derivatives), have become associated today with technological applications such as liquid-crystal displays, yet the importance of liquid crystals in the formation of biological structures is becoming ever more clear \citep{Haeckel,WentworthThompson:1917p2440, Needham, Bernal1967,Bouligand,Neville,Cartwright2007,Cartwright2009_2}.  

\subsection{Self-organization and self-assembly}

\begin{figure}[tp]
\centering\includegraphics*[width=\textwidth,clip=true]{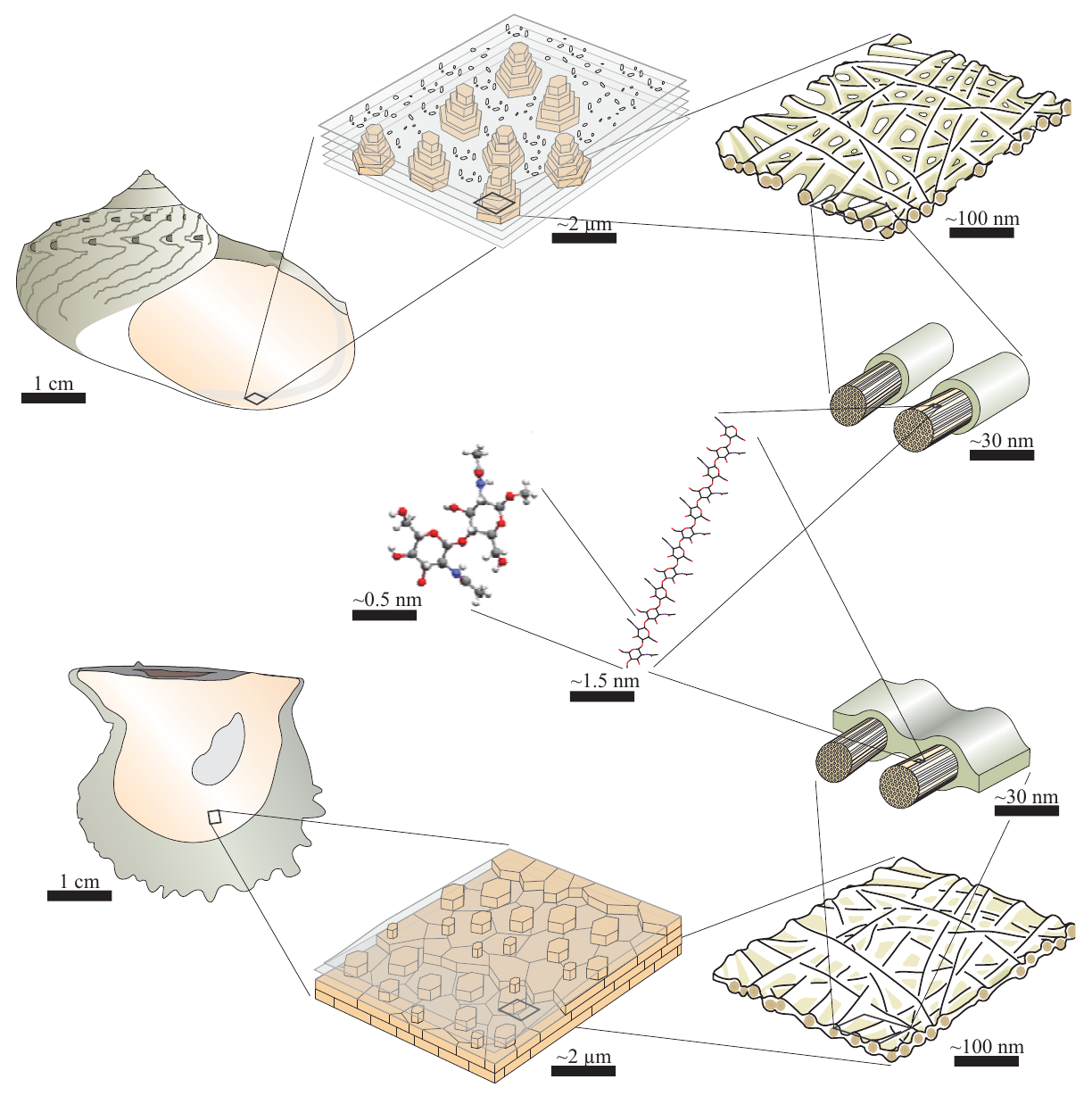} 
\caption{Hierarchization from the atomic to the macroscale in a biological system is exemplified in this diagram demonstrating nacre (mother of pearl) construction in gastropod and bivalve molluscs \citep{Cartwright2007}.
\label{nacre}}
\end{figure}

Liquid crystallization is just one physico-chemical process of self-organization and self-assembly that produces structures; in particular the structures of biological systems. There is increasing experimental evidence as to how biological and inorganic structures interact; how the biological system controls and exploits the inorganic world, and how structure is built upon structure leading to hierarchization from the atomic- to the macro-scale (see, for instance \cite{Whyte1968}); we give an example in Fig.~\ref{nacre}. Today there are efforts being made to understand the mathematical bases of different physical and chemical driving mechanisms of self-organization and self-assembly \citep{Ball:2001p3015,cademartiri} and their participation in biological processes \citep{WentworthThompson:1917p2440,Turing,Cartwright2009_1}. Among these mechanisms, we encounter solid and liquid crystallization alongside fluid flow, molecular diffusion and chemical reaction \citep{makki}. The attraction of understanding the mechanisms underlying pattern formation is clear: from the particular instance, one can extrapolate to an entire class of processes. To give just one example of the power of such a comprehensive understanding, excitability \citep{Ball:2001p3015} is a general mechanism that produces spiral and target patterns at various scales in a great variety of natural systems in physics, chemistry, biology and beyond, from the Belousov--Zhabotinsky reaction \citep{belousov1959,zhabotinsky1964_1,zhabotinsky1964_2,zaikin1970} to the myocardial tissue of the heart undergoing ventricular fibrillation. It turns out that the Burton--Cabrera--Frank mechanism of crystal growth \citep{Burton:1951p2792} is one more instance of an excitable medium \citep{Cartwright2012}; this mechanism of excitability functions in the self-assembly of crystals themselves, and hence the spiral and target patterns characteristic of excitable media appear at the molecular scale on the faces of growing crystals.

\subsection{The energy landscape}

One general concept that is proving useful to understand the basis for the particular structure of a crystal is that of an energy or structural landscape \citep{wales}. The potential energy function has local minima separated by energy barriers and
predictions can be made from knowledge of its stationary points, their topology and connectivity.  This energy-landscape approach provides a basis for crystal engineering \citep{tothadi}.  Protein folding is one instance in which considering the energy landscape approach proves helpful. In spite of the large number of metastable states, proteins fold (generally aided by solvent interactions and molecular chaperones) into their so-called native state --- essential for their biological functioning --- in a remarkably short time, and this may be understood by seeing the energy landscape as a funnel with the native state at the bottom. From the energy landscape one can grasp that polymorphism is the norm, rather than the exception, in crystallography, and make headway in the problem of calculating which polymorph will form in a given situation; a question depending not only on thermodynamics, but also on nucleation and growth kinetics. Likewise, amorphous solids or glasses can be viewed through the energy landscape approach; as with polymorphism, the existence of multiple amorphous states --- polyamorphism, the best-understood example of which is in ice \citep{bartels}.   --- is also not an exceptional state of affairs.

\subsection{Quasi- and aperiodic crystals}

\begin{figure}[tp]
\centering\includegraphics*[width=0.8\textwidth,clip=true]{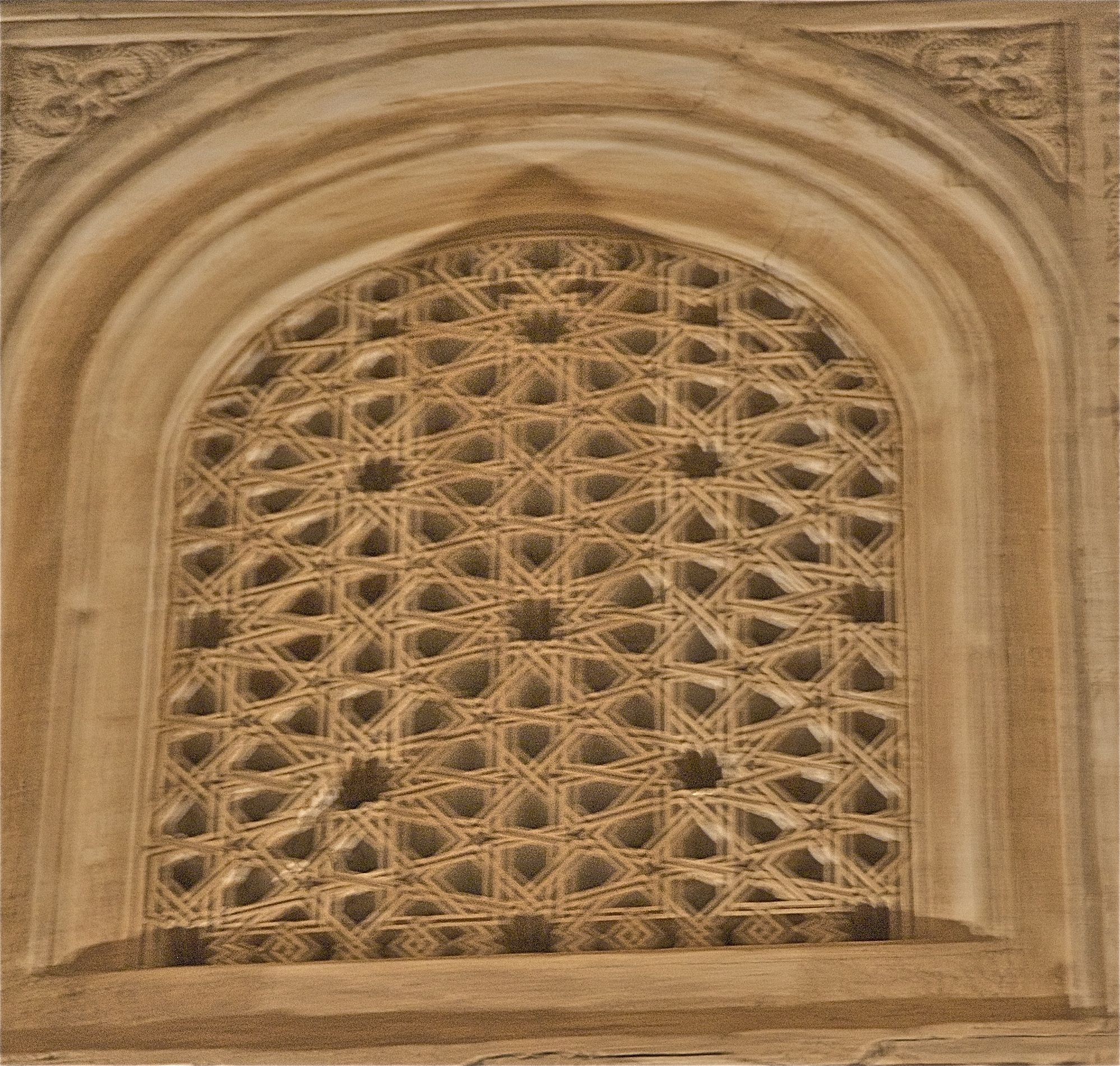} 
\caption{Decagonal patterns in a window of the Mirador de Lindaraja in the Alhambra, Granada.
\label{alhambra1}}
\end{figure}

The IUCr definition of a crystal that we discussed at the start was made to include the then newly-discovered quasicrystals. Understanding of these quasiperiodic structures, which have just given their discoverer Shechtman \citep{Shechtman} the Nobel prize, came from theoretical work \citep{Kramer,Hargittai} including that of \cite{Penrose} on non-periodic tilings, which, it was realized, should, if found in nature, give rise to a fivefold-symmetric diffraction pattern \citep{alan,alan2}. Such diffraction patterns turned up from the material produced by Shechtman. Further back in time, those who made Moorish art  were
interested in using such `forbidden symmetries'  \citep{zaslavsky,cromwell}, as shown in Fig.~\ref{alhambra1}. 
Several of the contributions to this issue relate to quasicrystals, including discussions of quantum behaviour in the presence of a quasiperiodic potential landscape \citep{vishveshwara}, epitaxy on quasicrystalline surfaces \citep{mcgrath}, and structures of pseudo-decagonal approximants in the Al-Co-Ni system \citep{hovmoller}. 
One can view quasicrystals as being intermediate between periodic crystals and amorphous structures, just as, in dynamical systems in the time domain, quasiperiodicity is intermediate between periodic behaviour and chaos. Of course, in the spatial domain, one has three dimensions to play with; quasicrystals can be quasiperiodic in one, two or three of these dimensions  \citep{Macia:2006tc}. 
One mostly open question is how to fabricate quasicrystals, especially in other materials apart from alloys. We should also note that these categories of periodic, quasiperiodic, and chaotic or amorphous are not limited to solid crystals; for example, liquid quasicrystals have also been described \citep{Zeng:2004vc,Zeng:2005dn},

%\subsection{Aperiodic crystals}

Likewise one can consider aperiodicity --- chaos --- in one, two or three dimensions. Amorphous solids or glasses are an example in three dimensions, but the amorphous structures known at present seem all to be, at best, metastable. Are there aperiodic crystals that are ground, equilibrium states; as \cite{Ruelle:1982hd} put it, `do turbulent crystals exist?' (Ruelle's paper is notable because it is written before the discovery of quasicrystals, but, based on the mathematics mentioned above, it takes for granted their existence: it begins ``we discuss the possibility that, besides periodic and quasiperiodic crystals, there exist turbulent crystals as thermodynamic equilibrium states at non-zero temperature''). In the introduction we mentioned silicon carbide polytypism; can this be viewed as one-dimensional amorphism?  For another example, cubic ice shows random stacking of cubic and hexagonal layers \citep{bartels}. Do these stacking faults imply that cubic ice might be viewed as a one-dimensional chaotic crystal? And might some small change in the initial conditions allow one to obtain a quasiperiodic arrangement of the layers; a one-dimensional quasicrystal of cubic ice? This intriguing thought brings us almost full circle to where we began with quasicrystals, in speculations on five-cornered snowflakes \citep{alan}.

\subsection{DNA and biological structure}

DNA is the molecule of life in which structure and information are intimately related. Before its structure was decoded by  \cite{watson}, it was noted by \cite{Bernal1931} that to be a practical information carrying system it should be a one-dimensional code; what \cite{Schrodinger} later called an aperiodic crystal. Bernal wrote:
\begin{quote}
``A complete molecule can be duplicated in three ways. If it is solid and three dimensional only a supernatural agency, a divine copyist, can, entering its inner complexity, reproduce it in detail. If we prefer a natural solution, we must imagine the molecule stretched out either in a plane or along a line. In either case the simpler constituent molecules have only to arrange themselves one by one on their identical partners in the original molecule, and then become linked to each other  by the absorption of suitable quanta from radiation or from second order collisions. That such autocatalysis is possible is indicated by recent work in Russia and America, where the regular atomic arrays of metallic catalysts are shown to operate like laceworkerÕs frames on which simple organic molecules settle to be joined into larger aggregates. A two-dimensional reproduction of this kind is impossible, owing to the fact that the constituent amino acids in nature are not symmetrical, but exist in right or left hand forms. Two-dimensional reproduction would lead to mirror image molecules, which are not found in nature. There remains then only one dimensional reproduction. At the moment of reproduction, but not necessarily at any other time, the molecule of the protein must be imagined as a pseudo-linear, associating itself, element by element, with identical groups, related by an axis instead of a plane of symmetry, and thus preserving only right or only left handed symmetry.''
\end{quote}
DNA represents the ultimate in information and structure: information is encoded in the DNA molecule through the genetic code.  Contributions to this issue address different aspects of DNA structure and information, from the mechanics of DNA \citep{Travers} to surprising mathematical symmetry features found in the genetic code \citep{Giannerini}.  
We now know that the ribosome is the molecular machine that reads information from (RNA) structure. To understand the dynamics of the ribosome, how it synchronizes itself with the correct `reading frame',
and the error correction features of the code,  ideas of the energy landscape will surely prove important.
Moreover, the algorithmic complexity of the human genome --- despite the relatively small number of genes it contains \citep{Claverie16022001}; the latest estimates indicate that there are some 23\,000 genes in the human genome--- is sufficient to construct a human being. How does so little information control so much behaviour? It is clear that genes must often act as choreographers, coding the big picture while leaving the detailed steps to be self-organized and self-assembled by physical and chemical processes.

\subsection{Revisiting Kepler's facultas formatrix}

%\section{Revisiting Kepler's facultas formatrix: The dynamics and informatics of self-organization and self-assembly}

Life itself must have emerged from self-organization and self-assembly \citep{Bernal1967}.
Vital steps at the origin of life must have involved events such as chiral symmetry breaking  \citep{Bonner1991,PhysRevLett.93.035502,PhysRevLett.98.165501} leading to the self-organization of the homochirality noted by Bernal, and the formation of membranes in the first proto-cells, for which ideas of the necessary self-assembly have been put forward both in the case of a hot origin of life \citep{mcglynn}, and in the case of a cold one \citep{bartels}. After millions of years of symbiosis and evolution from those first  proto-cells, today's cells have evolved extremely intricate interactions that control cell signalling and regulation whose workings may also be investigated in terms of the organization, assembly, and function of complex multicomponent systems \citep{bolanos}. 
A key question is how this increase in biological complexity throughout the evolutionary history of life has come about (e.g., \cite{Adami2000,PhysRevLett.84.6130}); must one posit a facultas formatrix?

In the 1960s \cite{whyte} pointed out that  a comprehensive scientific substitute was still lacking for the facultas formatrix, the self-organizing principle \cite{Kepler} formulated four hundred years ago while pondering the structure of the snowflake. Beyond snowflakes, the example of self-organization \emph{par excellence} must be life. If we consider life's origins, the appearance of self-organization and self-assembly and then self-replication of molecules has led to natural selection and thus evolution. The present understanding of self-organization and self-assembly, principally from work on dynamical systems and complexity, has done away with the facultas formatrix as some unfathomable formative capacity, as relativity swept away the luminiferous aether. The evolutionary history of life on Earth can be seen as a random walk in complexity space through which, from those initial proto-cells in their `warm little pond', or wherever, have evolved creatures with sufficient algorithmic complexity to ask these questions.

\section{Beyond crystals: A definition of structure based on the algorithmic complexity of its assembly}

Let us consider how to integrate information theory together with dynamical systems into this picture. (A related answer to this question, which may supplement or supplant our ideas, is put forward in a new review \citep{Crutchfield2012}.)
We commence by asking whether we may categorize structures based upon their complexity, derived from the ideas of algorithmic complexity of Kolmogorov and Chaitin \citep{kolmogorov1965,chaitin1966,li1997}. Given a structure, we might ask: What is the shortest algorithm that will describe the observed complexity? 
%Let us commence by asking whether we may regard a crystal as a structure, the description of which is very much smaller than the structure itself. This might seem to be the most general definition of a (perfect and infinite) crystal, derived from the ideas of algorithmic complexity of Kolmogorov and Chaitin \citep{kolmogorov1965,chaitin1966,li1997}. From this beginning let us generalize: let us consider categorizing structures based upon their complexity. Given a structure, we can ask: What is the shortest algorithm that will describe the observed complexity? 

The trouble with making use of the concept of Kolmogorov complexity in any application is that it is not a computable quantity; given a set of data one cannot, in general, obtain its Kolmogorov complexity. One cannot compute the information content of a given sequence because it is not possible in general to go backwards from the sequence to the algorithm that produced it. The nature of the problem, which is  a version of the Berry paradox, and linked to G\"odel's incompleteness theorem \citep{chaitin,hofstadter}, becomes clear when one reasons that a sequence of 100 digits may appear totally random --- and so possess maximum Kolmogorov complexity --- but in fact be 100 consecutive digits taken from, say, $\pi$, and so possess considerably less complexity. 

A further question concerning the algorithmic complexity of a physical structure is: at what level of description? The algorithmic complexity is expressible for an object encoded as a string of numbers, but that string expressing the description of a system and its complexity depends on the level in the hierarchy at which we choose to look at it. The level at which crystallographers describe a crystal is the atomic one, both above and below which there are other hierarchical levels. There is also the mathematical versus the physical description of crystal: one based on space groups and geometry compared to that based on energy minima and physics. On a human level we often describe a structure rather compactly e.g., ``a salt crystal''; even very complex ones: ``a human''.

The question of the uncomputability of Kolmogorov complexity is equivalent in a crystal (or other material) to recognizing that given a structure we cannot in general say how it was constructed. We know that in some systems there exist unconstructable structures, for example the so-called `garden of Eden' states of cellular automata, which cannot be reached from within the system. However, if a structure physically exists in our world, its constructibility is given, and moreover we immediately have in our hands a solution to this problem of computing the complexity, if one may consider the construction; the crystal growth; the assembly process that formed it. In other words, the same algorithm that builds a structure provides us with its complexity. We may add the corollary that if the possible structure corresponds to a point in an energy landscape, then the complexity of a particular structure comprises the route through the energy landscape.

We propose to call this measure the assembly complexity, and we should note that it is not necessarily coincident with the Kolmogorov complexity; in fact the two will only be the same if the assembly algorithm is the shortest possible, and since Kolmogorov complexity is not computable, this coincidence cannot in general be checked. In the case of a crystal, the shortest algorithm would simply be to repeat the unit cell, but assembly complexity would instead consider the crystal growth mechanism, such as the Burton--Cabrera--Frank mechanism we discussed above.

On what level of system description should we consider the assembly complexity? The level for a biological structure should clearly be that on which biological information is stored: the genomic one; let us term it the genomic complexity.  Again, the genomic complexity is not necessarily minimal, and so not necessarily coincident with Kolmogorov complexity; that would depend on the efficiency of the genome. 
Biology takes for granted physical and chemistry, so the level of description given in the genome is one that takes advantage of the physics and chemistry of nucleation, growth dynamics, fluid dynamics, solid and liquid crystallization, and so of self-organization and self-assembly.
The genome thus reflects the complexity of the machinery necessary to build the ingredients to assemble the structure: it is rather like describing food in terms of recipes. Of course environmental (epigenetic) factors also intervene in biological assembly (e.g., organisms must eat; ingest nutrients), just as we need to shop for the ingredients to make a recipe.

\begin{figure}[tp]
\centering\includegraphics*[angle=90,width=\textwidth,clip=true]{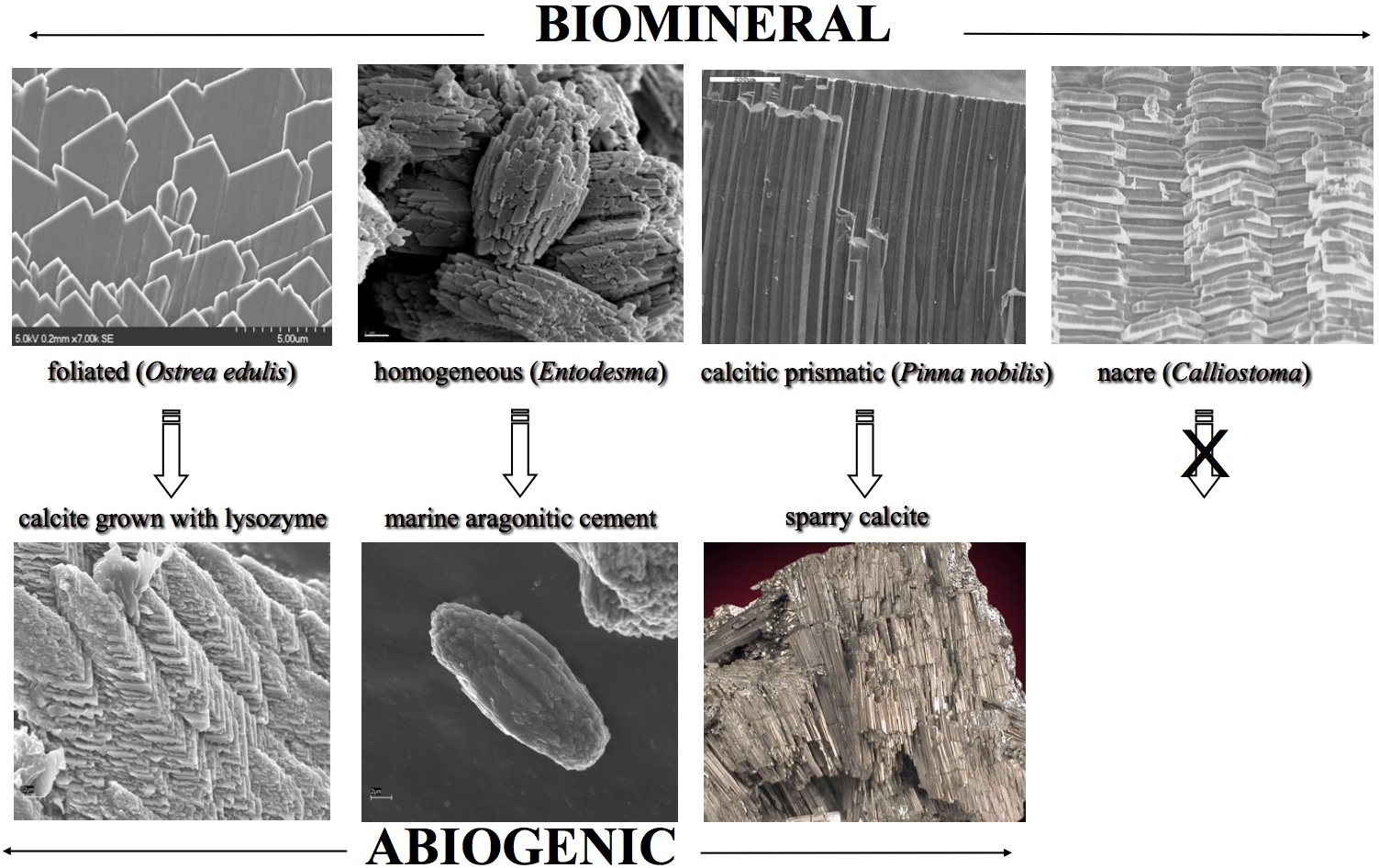} 
\caption{Some biomineral microstructures can be assimilated to abiogenic precipitates; not, however, nacre, with its characteristic brick-wall construction.
\label{biominerals}}
\end{figure}

Some biological materials can have low genomic complexity: as we noted above, in some instances genes leave considerable leeway for inorganic material to do its own thing. Good examples are to be found in biomineralization. The first three examples of biomineral microstructures in Fig.~\ref{biominerals} appear to be of this type, as we can see from their similarities to abiogenic precipitates (although it should be noted that the resemblance does not guarantee that they were assembled in the same fashion as the inorganic version, and until their genomics has been worked out it is only a hypothesis that they are of low genomic complexity). But consider the last example of a molluscan biomineral microstructure, nacre (mother of pearl). Its brick-wall structure has no resemblance to any inorganic structure. 
The algorithmic complexity of nacre is far greater than that of the previous biominerals, and presumably its genomic complexity, likewise. 

We illustrated in Fig.~\ref{nacre} how nacre is built up in a hierarchical fashion from a monomer of the polysaccharide chitin to the whole shell: chitin monomers are polymerized, and the polymer chains crystallized into rod-shaped crystallites; in turn these crystallites liquid crystallize into layers and are coated in proteins to become membranes; these membranes comprise the mortar of the brick wall. The space between the membranes, containing more proteins in solution, is then mineralized by tablets --- the bricks of the wall --- growing to fill the available space. The complex system that is nacre relies on a network of finely tuned interconnected physical and chemical mechanisms of the types we have discussed above, including solid and liquid crystallization, membrane dynamics, fluid dynamics, and so on, that must work in harmony to produce nacre.  Such hierarchized biological materials as nacre must have high genomic complexity that must be reflected in the portion of the genome that codes for nacre production.  
Being isolated, remote from the cells that secrete its components, it makes sense to think holistically of the set of constituents involved in mollusc shell biomineralization --- mineral, protein, polysaccharide, etc --- by analogy with the genome, proteome, and so on (the idea is due to Fr\'ed\'eric Marin). This specialized secretome might be termed the conchome, and the study of this complex system, conchomics.  (One might go further, and conceive of the complexity landscape of the conchome.) Nacre is just one product of the conchome, which encompasses nacre and the other microstructures of Fig.~\ref{biominerals}.
What are the genes for nacre? One question is how to isolate the part of the genome that corresponds to the conchome; progress is being made \cite{Jackson}. 
The important aspect for our purposes is that genomic complexity in biological systems links genetic information and structure.  

What do the genes for nacre do? They exercise tight control over the environment, and the nature and quantities of the ingredients.  That is the difference in biological materials. As \cite{mackay1986} stated:
``Taking a mineral comparable in complexity to a protein molecule we might ask, where are the genes in paulingite? We will anticipate the answer and reply that there are no genes, the information is distributed and is not localised. The rules of assembly of the whole emerge from the local rules, which in turn emerge from the energy levels of the individual atoms.''
For an abiotic system the environment --- provided by us, in the case of synthetic materials --- acts as the genes; in other words, we provide the secretome.  Just like in the biological case an abiotic system utilizes physics and chemistry leading to self-organization and self-assembly. In both biological and abiotic systems the important quantity in terms of the assembly complexity is how much and how often we must interact with the system to get it to behave as we wish it to, to produce the structure. Each time genes, or we, interact, that necessitates more complexity in the system. A given structure, even a complicated one like paulingite, might be totally `hands off', while the production of nacre requires a much more `hands on' approach even though individual parts of the hierarchical structure build themselves. 
Synthetic materials constructed up to now are of relatively low assembly complexity compared to natural materials that we cannot yet reproduce like nacre, wood, and silk. The intersection of the goals of biomimetics and the nanotechnology proposed by \cite{feynman} is to learn to imitate nature in this fashion to be able to produce artificial nacre, wood, silk and so on.
 
As long as we know how to build a structure, we can calculate its information content, from the simple algorithms such as the Burton--Cabrera--Frank mechanism of crystal growth to the complex hierarchized biological structures such as nacre through to the complex dynamics of the ribosome as a molecular machine that constructs proteins based on RNA plans copied from DNA. We propose that understanding in an algorithmic sense the very processes of self-organization and self-assembly resolves the dialectic \citep{alan1999} of integrating information theory into a new science of shape beyond crystals.

\begin{acknowledgements}
We thank Sho Takahashi for kindly allowing us to reproduce his artworks in Fig.~\ref{takahashi1} and on the cover of this issue. JHEC thanks Antonio Checa for Figs~\ref{sea_urchin1}, \ref{nacre}, and \ref{biominerals} and for biomineral discussions, John Finney for discussions on ice, Diego Gonz\'alez for a critical reading of the manuscript, and Fr\'ed\'eric Marin and Ana Vasiliu for discussions on nacre, and acknowledges MINCINN (Spain) project FIS2010-22322-528C02-02.
\end{acknowledgements}

\bibliographystyle{rspublicnat} 
\bibliography{beyond_crystals}

\end{document}